\pdfoutput=1

\documentclass[10pt,conference,review]{IEEEtran}
\IEEEoverridecommandlockouts
   
\usepackage{amsmath,amssymb,amsfonts}
\usepackage{graphicx}
\usepackage{textcomp}
\usepackage{xcolor}
\usepackage{cite}
\usepackage{adjustbox}
\usepackage{color, colortbl}
\usepackage{calc}
\usepackage{balance}

\usepackage{subfigure}

\usepackage{amsmath}
\usepackage{pgfplots}

\pgfplotsset{compat=1.10}
\usepgfplotslibrary{fillbetween}
\def\BibTeX{{\rm B\kern-.05em{\sc i\kern-.025em b}\kern-.08emT\kern-.1667em\lower.7ex\hbox{E}\kern-.125emX}}

\usepackage{array}    
\usepackage{booktabs} 
\usepackage[skins]{tcolorbox}
\usepackage{tabularx}
\usepackage{multirow}
\usepackage[flushleft]{threeparttable}
\usepackage{amssymb}
\usepackage{enumitem}
\usepackage{rotate}
\usepackage{pgf}
\usepackage{graphicx}
\usepackage{colortbl}
\usepackage{arydshln}
\usepackage{dblfloatfix} 
\usepackage{times}
\usepackage{rotating}
\usepackage{makecell} 
\usepackage{tabularx}
\usepackage{balance}
\usepackage{booktabs}
\usepackage{wrapfig}
\usepackage{tikz}
\usetikzlibrary{angles}
\usepackage{makecell}
\usepackage{tabu}
\newcommand{\bi}{\begin{itemize}}
\newcommand{\ei}{\end{itemize}}
\newcommand{\be}{\begin{enumerate}}
\newcommand{\ee}{\end{enumerate}}

\newcommand{\tbl}[1]{Table~\ref{tbl:#1}}

\newcommand{\tion}[1]{\S\ref{tion:#1}}
\usepackage{amsmath}
\usepackage{subfigure}

\usepackage{url}

\usepackage{tikz}
\usetikzlibrary{arrows.meta}
\tikzset{%
  >={Latex[width=2mm,length=2mm]},
            base/.style = {rectangle, rounded corners, draw=black,
                           minimum width=2.5cm, minimum height=1cm,
                           text centered, font=\sffamily},
  activityStarts/.style = {base, fill=blue!30},
       startstop/.style = {base, fill=red!30},
    activityRuns/.style = {base, fill=green!30},
         process/.style = {base, minimum width=2.5cm, fill=orange!15,
                           font=\ttfamily},
}

\makeatletter
\def\BState{\State\hskip-\ALG@thistlm}
\makeatother

\newcommand\MyBox[2]{
  \fbox{\lower0.75cm
    \vbox to 1.7cm{\vfil
      \hbox to 1.7cm{\hfil\parbox{1.4cm}{#1\\#2}\hfil}
      \vfil}%
  }%
}

\usepackage[final]{listings}
\usepackage{algorithm}
\usepackage[noend]{algpseudocode}

\usepackage[framemethod=tikz]{mdframed}
\usetikzlibrary{shadows}
\usepackage{graphics}
\newmdenv[
tikzsetting= {fill=gray!10},
linewidth=1pt,
roundcorner=2pt, 
shadow=false
]{myshadowbox}

\begin{document}

\title{Better Security Bug Report Classification \\via Hyperparameter Optimization}


\author{\IEEEauthorblockN{Rui Shu, Tianpei Xia, Laurie Williams, Tim Menzies}
\IEEEauthorblockA{Computer Science \\
North Carolina State University\\
Raleigh, USA \\
rshu, txia4, lawilli3@ncsu.edu, timm@ieee.org}

}

\maketitle
\thispagestyle{plain}
\pagestyle{plain}

\begin{abstract}
When security bugs are detected, they should be (a)~discussed privately by security software engineers; and (b)~not mentioned to the general public until security patches are available. Software engineers usually report bugs to bug tracking system, and label them as security bug reports (SBRs) or not-security bug reports (NSBRs), while SBRs have a higher priority to be fixed before exploited by attackers than NSBRs. Yet suspected security bug reports are often publicly disclosed because the mislabelling issues ( i.e., mislabel security bug reports as not-security bug report). 

The goal of this paper is to aid software developers to better classify bug reports that identify security vulnerabilities as security bug reports through parameter tuning of learners and data pre-processor. 
Previous work has applied text analytics and machine learning learners to classify which reported bugs are security related. We improve on that work, as shown by our analysis of five open source projects. We apply hyperparameter optimization to (a)~the control parameters of a learner; and (b)~the data pre-processing methods that handle the case where the target class is a small fraction of all the data. We show that optimizing the pre-processor is more useful than optimizing the learners. We also show that improvements gained from our approach can be very large. For example, using the same data sets as recently analyzed by our baseline approach, we show that adjusting the data pre-processing results in improvements to classification recall of 35\% to 65\% (median to max) with moderate increment of false positive rate. 

\end{abstract}

\begin{IEEEkeywords}
Hyperparameter Optimization, Data imbalance, Security Bug Report Classification
\end{IEEEkeywords}

\section{Introduction}

This paper replicates, and improves, the FARSEC a security bug report classification method recently reported in TSE'18~\cite{peters2018text}.
FARSEC found security bugs in five data sets with median recalls of 15,33,23,17,38\%. Our methods, applied to the same data, achieve much larger recalls of 77,66,67,44,62\%.
These better results were obtained after applying hyperparameter optimization to the data pre-processing. 
Such hyperparameter optimization has been used before in SE (e.g. for software defect classification~\cite{fu2016tuning,agrawal2018better} or effort estimation~\cite{xia2018hyperparameter}).  However, to the best of our knowledge, this is the first paper to apply
hyperparameter optimization to security bug detection.

We explore security bug detection since, to say the least, this a pressing current concern. Daily news reports reveal increasingly sophisticated security breaches.
As seen in those reports, a single vulnerability can have devastating effects. For example, a data breach of Equifax caused the personal information of as many as 143 million Americans -- or nearly half the country -- to be compromised~\cite{Equifax}. The WannaCry ransomware attack~\cite{WannaCry} crippled British medical emergency rooms, delaying medical procedures for many patients. 





Developers capture and document software bugs and issues in bug reports which are submitted to bug tracking systems. 
For example, the Mozilla bug database has maintained more than 670,000 bug reports with 135 new bug reports added each day~\cite{chen2013r2fix}. Usually, when security bugs are detected, they should be (a) discussed privately by security software engineers; and (b) not mentioned to the general public until security patches are available. Software engineering report bugs to bug tracking systems. Submitted bug report is explicitly labeled as a security bug report (SBR) or not-security bug report (NSBR). 


It is important to find and fix SBRs   before they can be exploited. To that end, researchers have adopted techniques, such as text mining, aiming at classifying bug reports accurately~\cite{gegick2010identifying,goseva2018identification,xia2014automated,xia2016automated}. One challenge faced by these techniques is that some security bug reports are mislabelled as not-security bug reports (which mainly results from a lack of security domain knowledge, e.g., inappropriately using security cross words in both security bug reports and not-security bug reports).



To address the mislabelling issue, several text mining based  classification models are proposed to distinguish security bug reports from not-security bug reports~\cite{gegick2010identifying}~\cite{goseva2018identification}. These approaches mainly identified relevant keywords in bug reports as well as features such as word frequency, which were then used in building classification models. However, these models were found to suffer from high false positive rate although they improved bug report classification. Peters et al. further proposed a text mining with irrelevancy pruning framework named FARSEC~\cite{peters2018text}. In their approach, developers first identified security related words. Next, they pruned away the irrelevant bug reports (where ``irrelevant'' meant ``does not have these security-related keywords'').
FARSEC was evaluated using bug reports from Chromium and four Apaches projects.

The conjecture of this paper is that text mining-based bug report classification approach (e.g., FARSEC) can be further enhanced. For example, FARSEC applied its data miners using their default ``off-the-shelf'' configurations.
We propose to apply hyperparameter optimization to the security bug report classification problem. Software engineering research results show that {\em hyperparameter optimization} (to automatically learn the ``magic'' control parameters of an algorithm) can outperform ``off-the-shelf'' configurations~\cite{agrawal2018wrong,agrawal2018better,fu2016tuning,herodotou2011starfish,tantithamthavorn2016automated,van2017automatic,krishna2018bad}. We distinguish and apply two kinds of hyperparamater optimization:

\bi
\item {\em Learner} hyperparamter optimization to adjust the parameters of the data miner; e.g. to learn how many trees to use in random forest, or what values to use in the kernel of a Support Vector Machine (SVM).
\item {\em Data pre-processor} hyperparamater optimization to adjust any adjustments to the training data, prior to learning; e.g. to learn how to control outlier removal or, in our case, how to handle the class imbalance problem. For example, data sets from ~\cite{peters2018text} show among the 45,940 bug reports, only 0.8\% belong to security bug reports.
\ei
This paper applies learner and pre-processor
optimization to vulnerability data sets as the state-of-the-art approach~\cite{peters2018text}. We conduct experiments on bug reports from five open source projects, i.e., one from the Chromium project, and four from Apache projects (i.e., Wicket, Ambari, Camel, Derby). We applied hyperparameter optimization to classification learners, and a hybrid sampling data pre-processor called SMOTE to address the class imbalance issue. We further propose to apply a tuned version of SMOTE called SMOTUNED in our experiment. Since FARSEC adopted several irrelevancy pruning techniques (i.e., filters) on data sets to remove not-security bug reports that contain security cross words, we also experimented on the same processed data as FARSEC.

We propose to ask the following research questions:

\newenvironment{RQ}{\vspace{2mm}\begin{tcolorbox}[enhanced,width=3.4in,size=fbox,fontupper=\small,colback=blue!5,drop shadow southwest,sharp corners]}{\end{tcolorbox}}
    
\begin{RQ}
{\bf RQ1.} Can learner hyperparameter optimization technique improve security bug report classification performance?
\end{RQ}

Using the same data as that of Peters et al. in the original FARSEC paper, we find that, hyperparameter optimization improves classification recall on 33 out of 40 filtering data sets than using the ``off-the-shelf'' default configurations in learners, with improvement from 8.6\% to 4.5X.

\begin{RQ}
{\bf RQ2.} Is it better to optimize the learners or the data pre-processors in security bug report classification?
\end{RQ}

The results from {\bf RQ1} were then compared to SMOTE and SMOTUNED (i.e., applying hyperparameter optimization to SMOTE). The improvements were seen with SMOTE and SMOTUNED were larger than with learner's hyperparamater optimization. Further, those observed improvements were very large indeed. Using the same data and text mining filtering methods as Peters et al., we can show improvements to recall of 35\% to 65\%, median to max
(at the cost of very modest increases in false alarm rates) 
 




\begin{RQ}
{\bf RQ3.} What are the relative merits of irrelevancy pruning (e.g., filtering) vs hyperparamater optimization in security bug report classification?
\end{RQ}

Our results show that {\em with data pre-processing optimization}, there is no added benefit to FARSEC's irrelevancy pruning. However, in a result confirming the Peters et al. results, {\em without hyperparamater optimization}, irrelevancy pruning does indeed improve the performance of vulnerability classification. So while we recommend data pre-processing optimization, if analysts wish to avoid that added complexity, they should use FARSEC's irrelevancy pruning.



The remainder of this paper is organized as follows. We introduce research background in Section~\ref{background}. In Section~\ref{evaluation}, we describe our method, including data, evaluation metrics and experiment details. We answer proposed research questions in section~\ref{results}. We discuss the threats to validity in Section~\ref{threats}. We then present related work in Section~\ref{related} and conclude in Section~\ref{conclusion}.


\section{Background}
\label{background}

\subsection{On the Need for More Secure Software}

As noted by the US National Institute of Standards and Technology (NIST), ``Current systems perform increasingly vital tasks and are widely known to possess vulnerabilities''~\cite{black2016dramatically}. Here, by a ``vulnerability'', they mean a weakness in the computational logic (e.g., code) found in software and some hardware components (e.g., firmware) that, when exploited, results in a negative impact on confidentiality, integrity, or availability~\cite{lewiscommon}. 

In their report to the White House Office of Science and Technology Policy, ``Dramatically Reducing Software Vulnerabilities''~\cite{black2016dramatically}, NIST officials urge the research community to explore more technical approaches to reducing security vulnerabilities. The need to reduce vulnerabilities is also emphasized in the 2016 US Federal Cybersecurity Research and Development Strategic Plan~\cite{strategicplan}. 

Experience shows that predicting for security vulnerabilities is difficult (see \tion{vuln}). Accordingly, this paper explores text-mining on bug reports, augmented by hyperparameter optimization and oversampling methods.

\subsection{FARSEC: Extending Text Mining for Bug  Reports}

\begin{table}[!b]
\caption {Different Filters used in FARSEC.}
\centering
\begin{tabular}{l|l}
\hline
\rowcolor[HTML]{EFEFEF} 
\multicolumn{1}{c|}{\cellcolor[HTML]{EFEFEF}\textbf{Filter}} & \multicolumn{1}{c}{\cellcolor[HTML]{EFEFEF}\textbf{Description}} \\ \hline
farsecsq & \begin{tabular}[c]{@{}l@{}}Apply the Jalali et al.~\cite{jalali2008optimizing} support function\\ to the frequency of words found in SBRs\end{tabular} \\ \hline
farsectwo & \begin{tabular}[c]{@{}l@{}}Apply the Graham version~\cite{graham2004hackers} of multiplying \\ the frequency by two.\end{tabular} \\ \hline
farsec & Apply no support function. \\ \hline
clni & Apply CLNI filter to non-filtered data. \\ \hline
clnifarsec & Apply CLNI filter to farsec filtered data. \\ \hline
clnifarsecsq & Apply CLNI filter to farsecsq filtered data. \\ \hline
clnifarsectwo & Apply CLNI filter to farsectwo filtered data. \\ \hline
\end{tabular}
\label{tbl:farsecFilter}
\end{table}

Recall that Peters et al. FARSEC's approach added an irrelevancy pruning step to text mining. \tbl{farsecFilter} lists the pruners explored in the FARSEC research. The purpose of filtering in FARSEC is to remove not-security bug reports with security related keywords. To achieve this goal, FARSEC applied an algorithm that firstly calculated the probability of the keyword appearing in security bug report and not-security bug report, and then calculated the score of the keyword.

Inspired by previous works~\cite{graham2004hackers}~\cite{jalali2008optimizing}, several tricks were also introduced in FARSEC to reduce false positives. For example, FARSEC built the {\em farsectwo} pruner by multiplying the frequency of not-security bug reports by two, aiming to achieve a good bias. The {\em farsecsq} filter was created by squaring the numerator of the support function to improve heuristic ranking of low frequency evidence.


In addition, FARSEC also tested a noise detection algorithm called CLNI (Closet List Noise Identification)~\cite{kim2011dealing}. Specifically, CLNI works as follows: During each iteration, for each instance $i$, a list of closest instances are calculated and sorted according to Euclidean Distance to instance $i$. The percentage of top $N$ instances with different class values is recorded. If percentage value is larger or equal to a threshold, then instance $i$ is highly probable to be a noisy instance and thus included to noise set $S$. This process is repeated until two noise sets $S_{i}$ and $S_{i-1}$ have the similarity over $\epsilon$. A threshold score (e.g., $0.75$) is set to remove any non-buggy reports above the score.

\subsection{Hyperparameter Optimization}\label{tion:hpo}

In machine learning, model parameters are the properties of training data that will learn on its own during training by the classifiers, e.g., split point in CART~\cite{breiman2017classification}. Model {\em hyperparameters} are values in machine learning models that can require different constraints, weights or learning rates to generate different data patterns, e.g., the number of neighbors in K Nearest Neighbors (KNN)~\cite{keller1985fuzzy}. 

Hyperparameters are important because they directly control the behaviors of the training algorithms and impact the performance of the models being trained. Choosing appropriate hyperparameters plays a critical role in the performance of machine learning models. {\em Hyperparameter optimization} is the process of searching the most optimal hyperparameters in machine learning learners~\cite{biedenkapp2018hyperparameter}~\cite{franceschi2017forward}. There are several kinds of hyperparameter optimization algorithms, including grid search, random search, and Bayesian optimization.

\textit{Grid search}~\cite{bergstra2011algorithms}~\cite{Tantithamthavorn:2016} is a ``brute force'' hyperparameter optimizer that wraps a learner in a  for-loops
that walk through a wide range of all a learner's control parameters. Simple to implement, it suffers from the ``curse of dimensionality''. That is, after just a  handful of options, grid search can miss important optimizations~\cite{fu2016differential}. Worse still, much CPU can be wasted during grid search since experience has shown that only a few ranges within a few tuning parameters really matter~\cite{bergstra2012random}.


\textit{Random search}~\cite{bergstra2012random} stochastically samples the search space and evaluates sets from a specified probability distribution. 
Evolutionary algorithms are a variant of random search that runs in ``generations''
where each new generation is seeded from the best examples selected from
the last generation~\cite{goldberg2006genetic}. Simulated annealing is a special form of evolutionary algorithms where the population size is one~\cite{kirkpatrick1983optimization}~\cite{Menzies:2007a}. The drawback of using random search algorithm is that it does not use information from prior experiment to select the next set and also it is very difficult to predict the next set of experiments.

\textit{Bayesian optimization}~\cite{pelikan1999boa} works by assuming the unknown function was sampled from a Gaussian Process and maintains a posterior distribution for this function as observation is made. Bayesian optimizers reflect on their distributions to propose the most informative guess about where to sample next. 
Such optimizers might be best-suited for optimization over continuous domains with a small number of dimensions~\cite{frazier2018tutorial}.

All the above algorithms have proven useful in their home domains. But for software engineering, certain algorithms such as differential evolution (DE)~\cite{storn1997differential} are known to run very quickly and deliver useful 
results~\cite{fu2016differential}~\cite{agrawal2018wrong}~\cite{agrawal2017better}. Also, DE has been proven useful in prior software engineering tuning studies~\cite{fu2016tuning}. Further, other evolutionary algorithms (e.g., genetic algorithms~\cite{goldberg2006genetic}, simulated annealing~\cite{kirkpatrick1983optimization}) mutate each attribute in isolation. When two attributes are correlated, those algorithms can mutate variables inappropriately in different directions. DE, on the other hand, mutates attributes in tandem along with known data trends. Hence, DE's tandem search can outperform other optimizers such as  
(a)~particle swarm optimization~\cite{vesterstrom2004comparative}; (b)~the grid search used by Tantithamthavorn et al.~\cite{tantithamthavorn2016automated}
to tune their defect predictors~\cite{fu2016differential}; or (c)~the genetic algorithm
used by Panichella et al. ~\cite{Panichella:2013} to tune a text miner.

The premise of DE is that the best way to mutate the existing tunings is to extrapolate between current solutions. Three solutions $a, b, c$ are selected at random. For each tuning parameter $k$, at some probability $cr$, we replace the old tuning $x_k$ with $y_k$. For booleans $y_k = \neg x_k$ and for numerics, \mbox{$y_k = a_k + f \times (b_k - c_k)$} where $f$ is a parameter controlling differential weight. DE loops $g$ times over the population of size $np$, replacing old items with new candidates (if new candidate is better). This means that, as the loop progresses, the population is full of increasingly more valuable solutions (which, in turn, helps extrapolation). As to the control parameters of DE, using advice from a DE user group\footnote{http://cci.lbl.gov/cctbx\_sources/scitbx/differential\_evolution.py},
we set $\{\mathit{np,f,cr}\}=\{10n,0.8,0.9\}$, where $n$ are the number of parameters to tune. Note that we set the number of iteration $\{\mathit{g}\}$ to $3$, $10$, which are denoted as DE$3$ and DE$10$ respectively. A small number ($3$) is used to test the effects of a very CPU-light effort estimator. A larger number ($10$) is selected to check if anything is lost by restricting the inference to small iterations.

\subsection{Data Balancing}

Many data sets
exhibit highly
imbalanced class frequencies (e.g., fake reviews in Amazon, fraudulent credit card charges, security bug reports). In these cases, only a small fraction of observations are actually positive because of the scarce occurrence of those events. In the field of data mining, such a phenomenon makes classification models difficult to detect rare events~\cite{sun2009classification}. Both misclassification and failing to identify rare events may result in poor performance by predictors.

One technique to tackle the imbalanced class issue is based on the sampling approach~\cite{haixiang2017learning}. There are three ways to resample imbalanced data sets~\cite{chawla2002smote,walden2014predicting,wallace2010semi,mani2003knn}: 

\begin{figure}[!t]
\centerline{\includegraphics[width=0.43\textwidth]{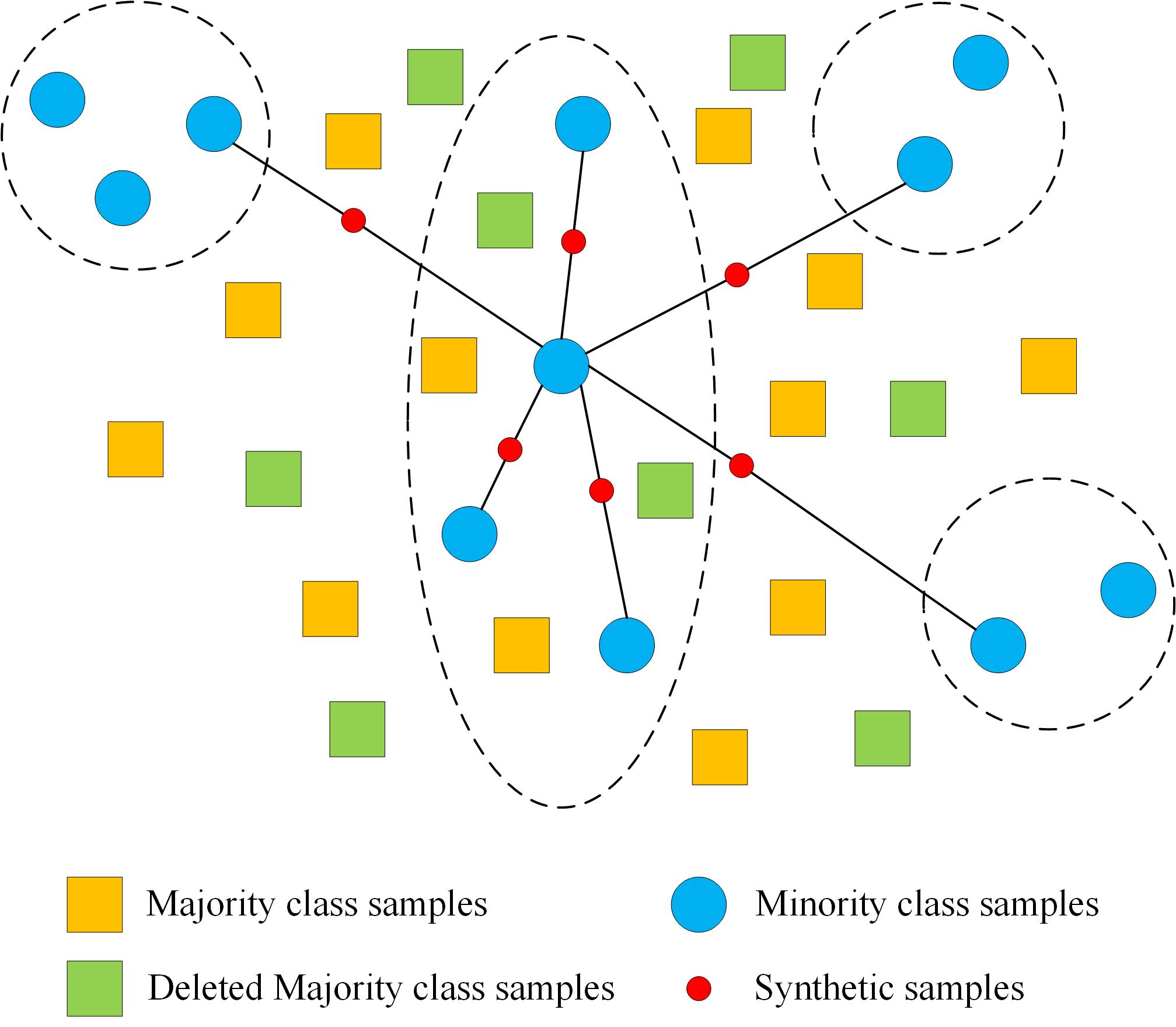}}
\caption{An illustration on how the SMOTE technique works.}    
\label{figure:SMOTE}
\end{figure}


\begin{table*}[!b]
\caption {List of hyperparameters optimized in different learners and pre-processor.}
\small
\resizebox{\textwidth}{!}{
\begin{tabular}{c|l|c|c|l}
\hline
\rowcolor[HTML]{EFEFEF}
\textbf{Learner \& Pre-processor} & \multicolumn{1}{c|}{\textbf{Parameters}} & \textbf{Default} & \textbf{Tuning Range} & \multicolumn{1}{c}{\textbf{Description}} \\ \hline
\multirow{6}{*}{Random Forest} & n\_estimators & 10 & {[}10,150{]} & The number of trees in the forest. \\ 
 & min\_samples\_leaf & 1 & {[}1,20{]} & The minimum number of samples required to be at a leaf node. \\  
 & min\_samples\_split & 2 & {[}2,20{]} & The minimum number of samples required to split an internal node. \\ 
 & max\_leaf\_nodes & None & {[}2,50{]} & Grow trees with  max\_leaf\_nodes in best-first fashion. \\  
 & max\_features & auto & {[}0.01,1{]} & The number of features to consider when looking for the best split \\  
 & max\_depth & None & {[}1,10{]} & The maximum depth of the tree. \\ \hline
\multirow{3}{*}{Logistic Regression} & C & 1.0 & {[}1.0,10.0{]} & Inverse of regularization strength. \\  
 & max\_iter & 100 & {[}50,200{]} & Useful only for the newton-cg, sag and lbfgs solvers. \\  
 & verbose & 0 & {[}0,10{]} & For the liblinear and lbfgs solvers set verbose to any positive number for verbosity. \\ \hline
\multirow{6}{*}{Multilayer Perceptron} & alpha & 0.0001 & {[}0.0001,0.001{]} & L2 penalty (regularization term) parameter. \\  
 & learning\_rate\_init & 0.001 & {[}0.001,0.01{]} & The initial learning rate used. \\
 & power\_t & 0.5 & {[}0.1,1{]} & The exponent for inverse scaling learning rate. \\
 & max\_iter & 200 & {[}50,300{]} & Maximum number of iterations. \\ 
 & momentum & 0.9 & {[}0.1,1{]} & Momentum for gradient descent update. \\ 
 & n\_iter\_no\_change & 10 & {[}1,100{]} & Maximum number of epochs to not meet tol improvement. \\ \hline
\multirow{2}{*}{K Nearest Neighbor} & leaf\_size & 30 & {[}10,100{]} & Leaf size passed to BallTree or KDTree. \\  
 & n\_neighbors & 5 & {[}1,10{]} & Number of neighbors to use by default for kneighbors queries. \\ \hline
Naive Bayes & var\_smoothing & 1e-9 & [0.0,1.0] & Portion of the largest variance of all features that is added to variances for calculation stability. \\ \hline
\multirow{3}{*}{SMOTE} & k & 5 & {[}1,20{]} & Number of neighbors. \\
 & m & 50\% & {[}50, 400{]} & \begin{tabular}[c]{@{}l@{}}Number of synthetic examples to create. Expressed as a percent of final training data.\end{tabular} \\
 & r & 2 & {[}1,6{]} & Power parameter for the Minkowski distance metric. \\ \hline
\end{tabular}
}
\label{table:hyperparameter}
\end{table*}

\bi
\item
Oversampling to make more of the minority class;
\item
Undersampling to remove majority class items; 
\item
Some hybrid of the first two.  
\ei

Machine learning researchers~\cite{haixiang2017learning} advise that undersampling can work better than oversampling if there are hundreds of minority observations in the datasets. When there are only a few dozen minority instances, oversampling approaches are superior to undersampling. In the case of large size of training samples, the hybrid methods would be a better choice.

The Synthetic Minority Oversampling TEchnique, also known as SMOTE~\cite{chawla2002smote}, is a hybrid algorithm that performs both over- and under-sampling. In oversampling, SMOTE calculates the $k$ nearest neighbors for each minority class samples. Depending on the amount of oversampling required, one or more of the $k$-nearest neighbors are picked to create the synthetic samples. This amount is usually denoted by oversampling percentage (e.g., $50\%$ by default). The next step is to randomly create a synthetic sample along the line connecting two minority samples. This can be illustrated in Figure~\ref{figure:SMOTE} where the blue dots represent the minority class sample, and the red dots represent the synthetic samples. In undersampling, SMOTE just removes majority samples randomly.

\begin{algorithm}[!t]
\small
\hspace{0.2cm}\begin{lstlisting}[xrightmargin=5.0ex,mathescape,frame=none]
def SMOTE(k=2, m=50%, r=2): # defaults
    while Majority > m do
        delete any majority sample
    while Minority < m do
        add synthetic samples
        
def synthetic_samples(X0):
    relevant = emptySet
    k1 = 0
    while(k1++ < 20) and size(relevant) < k {
        all = k1 nearest neighbors of X0
        relevant += item in all of X0 class
    }
    Z = any of relevant
    Y = interpolate(X0, Z)
    return Y
    
def minkowski_distance(a, b, r):
    return ($\Sigma_{i}abs(a_{i}-b_{i})^{r})^{1/r}$

\end{lstlisting}
\caption{Pseudocode of SMOTE From~\cite{agrawal2017better}.}\label{algorithm:smote}  
\end{algorithm}

Algorithm~\ref{algorithm:smote} shows the SMOTE algorithm and Table~\ref{table:hyperparameter} shows the ``magic'' parameters that control SMOTE. The best settings for these magic parameters are often domain-specific. SMOTUNED~\cite{agrawal2018better} is a tuned version of SMOTE that uses DE to learn good settings for the Table~\ref{table:hyperparameter} parameters. SMOTUNED will serve as our optimizer for data pre-processing to handle data imbalance for security bug classification.



Before leaving this description of SMOTE, we make one important methodological point.
Data mining algorithms should be assessed on the kinds of data they might see in the future. That is, data miners should be tested on data that has {\em naturally occurring class distributions}. Hence, while applying SMOTE (or SMOTUNED) to the {\em training data} is good practice, it is a bad practice to apply SMOTE to the {\em testing data}.

\subsection{Machine Learning Algorithms}
Once SMOTE, or SMOTUED terminates, the resulting data must be processed by a data miner. This study using the same five machine learning learners as seen in the FARSEC study, i.e., Random Forest, Naive Bayes, Logistic Regression, Multilayer Perceptron and K Nearest Neighbor. They are widely used for software engineering classification problems~\cite{lessmann2008benchmarking}. We now give a brief description of each algorithm.

\textbf{Random Forest.} Random forests~\cite{liaw2002classification} are an ensemble learning that combines multiple weaker learners into a stronger learner. Breiman argues that such random forests have better generalization are less susceptible to over-fitting~\cite{Breiman2001}. There are several advantages of random forest, e.g. it can be used for both classification and regression problems, and it is easy to measure the relative importance of each feature on the classification.

\textbf{Naive Bayes.} Naive Bayes~\cite{rish2001empirical} keep statistics on each column of data in each class. New examples are classified by a statistical analysis that reports the class that is closest to the test case. Naive Bayes classifiers are computationally efficient and tend to perform well on relatively small data sets.

\textbf{Logistic Regression.} Logistic regression~\cite{hosmer2013applied} is a classification model that works well on linearly separable classes, and it is one of the most widely used algorithms for classification in industry (e.g., weather forecasting). In general, logistic regression could be viewed as a probability estimation. A threshold can be set to predict the class a data belongs. Based upon the threshold, the obtained estimated probability is classified into classes. The decision boundary of logistic regression can be linear or non-linear. 

\textbf{Multilayer Perceptron.} Multilayer perceptron~\cite{gardner1998artificial} is an artificial neural network which is composed of multiple perceptrons. A perceptron is a single neuron model that is a precursor to larger neural networks. An input layer receives the signal and an output layer makes the classification about the input. Between the input layer and the output layer, an arbitrary number of hidden layers act as the computational engine. In a sense, multilayer perceptron learns a mapping of the input training data, and best relate it to the predicted output. 

\textbf{K Nearest Neighbor.} K nearest neighbor classifier~\cite{keller1985fuzzy} is one of the simplest predictive models. When classifying a new element, nearest neighbor classifier checks in the training data for K objects close to the element (i.e., K nearest neighbors). The classifier returns the class by majority voting among its neighbors. K nearest neighbor is a typical ``lazy learner'' since it does not learn a discriminative function from the training data but just memorizes the training data instead.

\subsection{Hyperparameter Tuning Range of Learners}\label{tion:range}

We implement our hyperparameter optimization techniques in scikit-learn~\cite{pedregosa2011scikit}. We choose scikit-learn because it integrates various well-defined classification learners as well as many configurations that we can choose. We select a group of hyperparameters for each learner in scikit-learn for the purpose of optimization. Table~\ref{table:hyperparameter} lists all the hyperparameters we select. We choose these ``tuning range'' for several reasons: (1) These ranges include the default value. (2) We could achieve a large gains in performance of our vulnerability predictors by exploring these ranges. (3) These ranges are sufficient in tuning for this work.

Besides, \tbl{DEpara} shows the DE control parameters used in this paper. This table was generated by combining the advice at the end of \tion{hpo} with Table~\ref{table:hyperparameter}.

\begin{table}[!t]
\caption {List of parameters in DE algorithm for different learners and pre-processor.}
\centering
\begin{tabular}{l|c|c|c|c}
\hline
\rowcolor[HTML]{EFEFEF} 
\multicolumn{1}{c|}{\cellcolor[HTML]{EFEFEF}} & \multicolumn{4}{c}{\cellcolor[HTML]{EFEFEF}\textbf{DE Parameter}} \\ \cline{2-5} 
\rowcolor[HTML]{EFEFEF} 
\multicolumn{1}{c|}{\multirow{-2}{*}{\cellcolor[HTML]{EFEFEF}\textbf{Learner \& Pre-processor}}} & \textbf{NP} & \textbf{F} & \textbf{CR} & \textbf{ITER} \\ \hline
Random Forest & 60 &  &  &  \\ \cline{1-2}
Logistic Regression & 30 &  &  &  \\ \cline{1-2}
Multilayer Perceptron & 60 &  &  &  \\ \cline{1-2}
K Nearest Neighbor & 20 &  &  &  \\ \cline{1-2}
Naive Bayes & 10 & \multirow{-5}{*}{0.8} & \multirow{-5}{*}{0.9} & \multirow{-5}{*}{3, 10} \\ \hline
SMOTE & 30 & 0.8 & 0.9 & 10 \\ \hline
\end{tabular}
\label{tbl:DEpara}
\end{table}

\section{Methods}
\label{evaluation}

\subsection{Experimental Rig}
The results of this paper come from a 10-way cross-validation. After dividing each {\em training data} into $B=10$ bins, we tested our models using bin $B_i$ after training them on {\em data $- B_i$}. The 10-way cross-validation is used to pick the best candidate learner with the highest performance for that data set. We then train the candidate learner with the whole training data set, and test on the separate testing data set as FARSEC.

This 10-way process was applied to five projects (described in the next section);
the five learners listed above; and the following eight data pruning operators:
\bi
\item {\em train}; i.e.  no data pruning;
\item The seven variations of  {\em farsec} and
{\em clni} listed in \tbl{farsecFilter}. 
\ei
That is, we ran our learners 2000 times: 
\[
 5\;{\mathit projects}  * 8\;{\mathit pruners} * 5\;{\mathit learners}  *  10\;{\mathit ways} 
\]


Note that FARSEC experimented on bug reports from five open source projects, i.e., one Chromium project and four Apache projects (i.e., Wicket, Ambari, Camel and Derby). Table~\ref{tbl:farsecDataset} shows the data imbalanced characteristics of the bug reports.

\begin{table}[!t]
\caption {Imbalanced characteristics of bug report data sets from FARSEC.}
\footnotesize
\centering
\begin{tabular}{c|l|c|r|r}
\hline
\rowcolor[HTML]{EFEFEF} 
  & \multicolumn{1}{c|}{~} & \textbf{\#Security} & \multicolumn{1}{c|}{\textbf{\#Bug}} & \multicolumn{1}{c}{{\bf Percent}} \\
\rowcolor[HTML]{EFEFEF} 
\textbf{Project} & 
\multicolumn{1}{c|}{\textbf{Filter}} &
\textbf{Bugs} & \multicolumn{1}{c|}{\textbf{reports}} & \multicolumn{1}{c}{{\bf buggy}} 
\\ \hline
\multirow{8}{*}{Chromium} & train & \multirow{8}{*}{77} & 20970 & 0.4 \\  
 & farsecsq &  & 14219 & 0.5 \\ 
 & farsectwo &  & 20968 & 0.4 \\  
 & farsec &  & 20969 & 0.4 \\  
 & clni &  & 20154 & 0.4 \\  
 & clnifarsecsq &  & 13705 & 0.6 \\  
 & clnifarsectwo &  & 20152 & 0.4 \\  
 & clnifarsec &  & 20153 & 0.4 \\ \hline
\multirow{8}{*}{Wicket} & train & \multirow{8}{*}{4} & 500 & 0.8 \\
 & farsecsq &  & 136 & 2.9 \\  
 & farsectwo &  & 143 & 2.8 \\  
 & farsec &  & 302 & 1.3 \\  
 & clni &  & 392 & 1.0 \\  
 & clnifarsecsq &  & 46 & 8.7 \\  
 & clnifarsectwo &  & 49 & 8.2 \\  
 & clnifarsec &  & 196 & 2.0 \\ \hline
\multirow{8}{*}{Ambari} & train & \multirow{8}{*}{22} & 500 & 4.4 \\  
 & farsecsq &  & 149 & 14.8 \\  
 & farsectwo &  & 260 & 8.5 \\  
 & farsec &  & 462 & 4.8 \\  
 & clni &  & 409 & 5.4 \\  
 & clnifarsecsq &  & 76 & 28.9 \\ 
 & clnifarsectwo &  & 181 & 12.2 \\  
 & clnifarsec &  & 376 & 5.9 \\ \hline
\multirow{8}{*}{Camel} & train & \multirow{8}{*}{14} & 500 & 2.8 \\  
 & farsecsq &  & 116 & 12.1 \\ 
 & farsectwo &  & 203 & 6.9 \\  
 & farsec &  & 470 & 3.0 \\  
 & clni &  & 440 & 3.2 \\  
 & clnifarsecsq &  & 71 & 19.7 \\  
 & clnifarsectwo &  & 151 & 9.3 \\  
 & clnifarsec &  & 410 & 3.4 \\ \hline
\multirow{8}{*}{Derby} & train & \multirow{8}{*}{46} & 500 & 9.2 \\  
 & farsecsq &  & 57 & 80.7 \\  
 & farsectwo &  & 185 & 24.9 \\  
 & farsec &  & 489 & 9.4 \\  
 & clni &  & 446 & 10.3 \\  
 & clnifarsecsq &  & 48 & 95.8 \\  
 & clnifarsectwo &  & 168 & 27.4 \\  
 & clnifarsec &  & 435 & 10.6 \\ \hline
\end{tabular}
\label{tbl:farsecDataset}
\end{table}

\subsection{Evaluation Metrics}

To understand the open issues with vulnerability classification, firstly we must define how they are \textbf{assessed}. If (TN, FN, FP, TP) are the true negatives, false negatives, false positives, and true positives, respectively, found by a detector, then: 

\bi
\item
{\em prec} = Precision = TP/(TP+FP), the percentage of the predicted vulnerabilities that are actual vulnerabilities; 
\item
{\em pd} = Recall = TP/(TP+FN), the percentage of the actual vulnerabilities that are predicted to be vulnerabilities; 
\item
{\em pf} = False Alarms = FP/(FP+TN), the percentage of the non-vulnerable artifacts that are reported as vulnerable;  
\ei

This paper adopts the same evaluation criteria of the original FARSEC paper; i.e. the recall ({\em pd}) and false alarm ({\em pf}) measures. Also, to control the optimization of differential evolution algorithm, we instructed it to minimize false alarms while maximizing recall. To achieve those goals, we told DE to maximize the {\em g-measure} which  is the harmonic mean of recall and the compliment of false alarms. 
\[
 g = \frac{2 \times \mathit{pd} \times (1-\mathit{pf})}{\mathit{pd} + (1-\mathit{pf})}
\]
Note that {\em g} is maximal when {\em both} recall ({\em pd}) is high and false alarm ({\em pf}) is low.




\begin{table*}[!htb] 
\caption {Recall and false positive results from FARSEC, default Scikit-learn and DE tuning with Scikit-learn., SMOTE with Scikit-learn and SMOTUNED with Scikit-learn.
In these results, {\em higher} recalls (a.k.a. pd)
are {\em better}. Also, {\em lower} false alarms
(a.k.a. pf) are {\em better}. Cells shown in gray report
a statistical comparison of all the recall (a.k.a. pd) results. In any row, all the ``best'' recall results are colored in gray
(and by ``best'', we mean those results that appear in the top-rank of a statistical Scott-Knot test for statistical
significance and non-trival effect size).
}

\vspace{8mm} 
\centering
\begin{tabular}{l|l|c|r|r|c|r|r|c|r|r|c|r|r}
\hline
\rowcolor[HTML]{EFEFEF} 
\multicolumn{1}{c|}{\cellcolor[HTML]{EFEFEF}} & \multicolumn{1}{c|}{\cellcolor[HTML]{EFEFEF}} & \multicolumn{3}{c|}{\cellcolor[HTML]{EFEFEF}\textbf{\begin{tabular}[c]{@{}c@{}}FARSEC\\ (``off-the-shelf'' WEKA)\end{tabular}}} & \multicolumn{3}{c|}{\cellcolor[HTML]{EFEFEF}\textbf{\begin{tabular}[c]{@{}c@{}}Learning Tuning\\ (Scikit-learn)\end{tabular}}} & \multicolumn{3}{c|}{\cellcolor[HTML]{EFEFEF}\textbf{\begin{tabular}[c]{@{}c@{}}SMOTE\\ (Scikit-learn)\end{tabular}}} & \multicolumn{3}{c}{\cellcolor[HTML]{EFEFEF}\textbf{\begin{tabular}[c]{@{}c@{}}SMOTUNED\\(Scikit-learn)\end{tabular}}} \\ \cline{3-14} 
\rowcolor[HTML]{EFEFEF} 
\multicolumn{1}{c|}{\multirow{-2}{*}{\cellcolor[HTML]{EFEFEF}\textbf{Project}}} & \multicolumn{1}{c|}{\multirow{-2}{*}{\cellcolor[HTML]{EFEFEF}\textbf{Filter}}} & \textbf{Learner} & \multicolumn{1}{c|}{\cellcolor[HTML]{EFEFEF}\textbf{pd}} & \multicolumn{1}{c|}{\cellcolor[HTML]{EFEFEF}\textbf{pf}} & \textbf{Learner} & \multicolumn{1}{c|}{\cellcolor[HTML]{EFEFEF}\textbf{pd}} & \multicolumn{1}{c|}{\cellcolor[HTML]{EFEFEF}\textbf{pf}} & \textbf{Learner} & \multicolumn{1}{c|}{\cellcolor[HTML]{EFEFEF}\textbf{pd}} & \multicolumn{1}{c|}{\cellcolor[HTML]{EFEFEF}\textbf{pf}} & \textbf{Learner} & \multicolumn{1}{c|}{\cellcolor[HTML]{EFEFEF}\textbf{pd}} & \multicolumn{1}{c}{\cellcolor[HTML]{EFEFEF}\textbf{pf}} \\ \hline
 & train & LR & 15.7 & 0.2 & NB & 46.9 & 6.8 & NB & 68.7 & 24.1 & \cellcolor[HTML]{EFEFEF}MP & \cellcolor[HTML]{EFEFEF}73.9 & 17.8 \\  
 & farsecsq & RF & 14.8 & 0.3 & LR & 64.3 & 10.3 & NB & 80.0 & 31.5 & \cellcolor[HTML]{EFEFEF}LR & \cellcolor[HTML]{EFEFEF}84.3 & 25.1 \\  
 & farsectwo & LR & 15.7 & 0.2 & NB & 40.9 & 6.5 & \cellcolor[HTML]{EFEFEF}RF & \cellcolor[HTML]{EFEFEF}78.3 & 27.6 & RF & 77.4 & 23.1 \\  
 & farsec & LR & 15.7 & 0.2 & NB & 46.1 & 6.9 & \cellcolor[HTML]{EFEFEF}RF & \cellcolor[HTML]{EFEFEF}80.8 & 36.1 & MP & 72.2 & 14.9 \\  
 & clni & LR & 15.7 & 0.2 & NB & 30.4 & 4.1 & \cellcolor[HTML]{EFEFEF}NB & \cellcolor[HTML]{EFEFEF}74.8 & 24.8 & MP & 72.2 & 13.6 \\  
 & clnifarsecsq & MP & 49.6 & 3.8 & MP & 72.2 & 14.2 & NB & 82.6 & 30.4 & \cellcolor[HTML]{EFEFEF}LR & \cellcolor[HTML]{EFEFEF}86.1 & 25.6 \\  
 & clnifarsectwo & LR & 15.7 & 0.2 & NB & 50.4 & 7.0 & \cellcolor[HTML]{EFEFEF}NB & \cellcolor[HTML]{EFEFEF}79.1 & 29.9 & MP & 74.8 & 12.8 \\  
\multirow{-8}{*}{Chromium} & clnifarsec & LR & 15.7 & 0.2 & NB & 47.8 & 10.4 & \cellcolor[HTML]{EFEFEF}NB & \cellcolor[HTML]{EFEFEF}78.3 & 29 & MP & 74.7 & 17.1 \\ 
\multicolumn{14}{r}{~}\\\hline
 & train & NB & 16.7 & 7.1 & NB & 0.0 & 5.1 & \cellcolor[HTML]{EFEFEF}NB & \cellcolor[HTML]{EFEFEF}66.7 & 32 & \cellcolor[HTML]{EFEFEF}NB & \cellcolor[HTML]{EFEFEF}66.7 & 12.1 \\  
 & farsecsq & LR & 66.7 & 38.3 & NB & 50.0 & 44.5 & \cellcolor[HTML]{EFEFEF}NB & \cellcolor[HTML]{EFEFEF}83.3 & 71.3 & \cellcolor[HTML]{EFEFEF}NB & \cellcolor[HTML]{EFEFEF}83.3 & 66.8 \\  
 & farsectwo & \cellcolor[HTML]{EFEFEF}LR & \cellcolor[HTML]{EFEFEF}66.7 & 36.6 & NB & 50.0 & 42.3 & \cellcolor[HTML]{EFEFEF}NB & \cellcolor[HTML]{EFEFEF}66.7 & 68.2 & \cellcolor[HTML]{EFEFEF}NB & \cellcolor[HTML]{EFEFEF}66.7 & 62.9 \\  
 & farsec & LR & 33.3 & 8.1 & \cellcolor[HTML]{EFEFEF}NB & \cellcolor[HTML]{EFEFEF}66.7 & 23.1 & \cellcolor[HTML]{EFEFEF}NB & \cellcolor[HTML]{EFEFEF}66.7 & 43.9 & \cellcolor[HTML]{EFEFEF}NB & \cellcolor[HTML]{EFEFEF}66.7 & 26.1 \\  
 & clni & NB & 0.0 & 5.5 & MP & 16.7 & 2.4 & \cellcolor[HTML]{EFEFEF}NB & \cellcolor[HTML]{EFEFEF}50.0 & 21.1 & \cellcolor[HTML]{EFEFEF}NB & \cellcolor[HTML]{EFEFEF}50.0 & 12.5 \\  
 & clnifarsecsq & LR & 33.3 & 25.5 & \cellcolor[HTML]{EFEFEF}NB & \cellcolor[HTML]{EFEFEF}83.3 & 66.8 & \cellcolor[HTML]{EFEFEF}NB & \cellcolor[HTML]{EFEFEF}83.3 & 66.8 & \cellcolor[HTML]{EFEFEF}NB & \cellcolor[HTML]{EFEFEF}83.3 & 66.8 \\  
 & clnifarsectwo & LR & 33.3 & 27.7 & NB & 50.0 & 39.9 & \cellcolor[HTML]{EFEFEF}NB & \cellcolor[HTML]{EFEFEF}66.7 & 61.3 & \cellcolor[HTML]{EFEFEF}NB & \cellcolor[HTML]{EFEFEF}66.7 & 61.3 \\  
\multirow{-8}{*}{Wicket} & clnifarsec & LR & 50.0 & 10.5 & \cellcolor[HTML]{EFEFEF}NB & \cellcolor[HTML]{EFEFEF}66.7 & 23.1 & \cellcolor[HTML]{EFEFEF}NB & \cellcolor[HTML]{EFEFEF}66.7 & 38.9 & \cellcolor[HTML]{EFEFEF}NB & \cellcolor[HTML]{EFEFEF}66.7 & 22.9 

\\ 
\multicolumn{14}{r}{~}\\\hline

 & train & MP & 14.3 & 1.6 & LR & 28.6 & 0.8 & \cellcolor[HTML]{EFEFEF}LR & \cellcolor[HTML]{EFEFEF}57.1 & 20.1 & \cellcolor[HTML]{EFEFEF}RF & \cellcolor[HTML]{EFEFEF}57.1 & 10.8 \\  
 & farsecsq & RF & 42.9 & 14.4 & \cellcolor[HTML]{EFEFEF}RF & \cellcolor[HTML]{EFEFEF}57.1 & 2.8 & \cellcolor[HTML]{EFEFEF}LR & \cellcolor[HTML]{EFEFEF}57.1 & 30.4 & \cellcolor[HTML]{EFEFEF}RF & \cellcolor[HTML]{EFEFEF}57.1 & 17.2 \\  
 & farsectwo & \cellcolor[HTML]{EFEFEF}RF & \cellcolor[HTML]{EFEFEF}57.1 & 3.0 & \cellcolor[HTML]{EFEFEF}RF & \cellcolor[HTML]{EFEFEF}57.1 & 2.8 & \cellcolor[HTML]{EFEFEF}RF & \cellcolor[HTML]{EFEFEF}57.1 & 22.1 & \cellcolor[HTML]{EFEFEF}RF & \cellcolor[HTML]{EFEFEF}57.1 & 17.8 \\  
 & farsec & MP & 14.3 & 4.9 & \cellcolor[HTML]{EFEFEF}RF & \cellcolor[HTML]{EFEFEF}57.1 & 2.0 & \cellcolor[HTML]{EFEFEF}LR & \cellcolor[HTML]{EFEFEF}57.1 & 19.9 & \cellcolor[HTML]{EFEFEF}NB & \cellcolor[HTML]{EFEFEF}57.1 & 7.1 \\  
 & clni & MP & 14.3 & 2.6 & LR & 28.6 & 0.8 & \cellcolor[HTML]{EFEFEF}LR & \cellcolor[HTML]{EFEFEF}57.1 & 12.4 & \cellcolor[HTML]{EFEFEF}LR & \cellcolor[HTML]{EFEFEF}57.1 & 8.9 \\  
 & clnifarsecsq & \cellcolor[HTML]{EFEFEF}RF & \cellcolor[HTML]{EFEFEF}57.1 & 7.7 & \cellcolor[HTML]{EFEFEF}RF & \cellcolor[HTML]{EFEFEF}57.1 & 2.4 & \cellcolor[HTML]{EFEFEF}RF & \cellcolor[HTML]{EFEFEF}57.1 & 13.4 & \cellcolor[HTML]{EFEFEF}RF & \cellcolor[HTML]{EFEFEF}57.1 & 7.1 \\  
 & clnifarsectwo & RF & 28.6 & 4.5 & \cellcolor[HTML]{EFEFEF}RF & \cellcolor[HTML]{EFEFEF}57.1 & 2.8 & \cellcolor[HTML]{EFEFEF}RF & \cellcolor[HTML]{EFEFEF}57.1 & 13.0 & \cellcolor[HTML]{EFEFEF}RF & \cellcolor[HTML]{EFEFEF}57.1 & 5.1 \\  
\multirow{-8}{*}{Ambari} & clnifarsec & RF & 14.3 & 0.0 & \cellcolor[HTML]{EFEFEF}RF & \cellcolor[HTML]{EFEFEF}57.1 & 2.4 & \cellcolor[HTML]{EFEFEF}RF & \cellcolor[HTML]{EFEFEF}57.1 & 7.9 & \cellcolor[HTML]{EFEFEF}RF & \cellcolor[HTML]{EFEFEF}57.1 & 3.9 \\ 

\\ 
\multicolumn{14}{r}{~}\\\hline

 & train & LR & 11.1 & 3.5 & MP & 16.7 & 1.5 & NB & 33.3 & 27.4 & \cellcolor[HTML]{EFEFEF}NB & \cellcolor[HTML]{EFEFEF}44.4 & 35.9 \\  
 & farsecsq & RF & 16.7 & 11.4 & RF & 44.4 & 24.7 & RF & 44.4 & 20.5 & \cellcolor[HTML]{EFEFEF}RF & \cellcolor[HTML]{EFEFEF}55.6 & 23.4 \\  
 & farsectwo & LR & 50.0 & 41.8 & RF & 44.4 & 17.6 & \cellcolor[HTML]{EFEFEF}NB & \cellcolor[HTML]{EFEFEF}61.1 & 71.0 & \cellcolor[HTML]{EFEFEF}NB & \cellcolor[HTML]{EFEFEF}61.1 & 53.1 \\  
 & farsec & LR & 16.7 & 6.9 & NB & 22.2 & 12.4 & \cellcolor[HTML]{EFEFEF}NB & \cellcolor[HTML]{EFEFEF}33.3 & 39.4 & \cellcolor[HTML]{EFEFEF}NB & \cellcolor[HTML]{EFEFEF}33.3 & 28.0 \\  
 & clni & NB & 16.7 & 12.3 & NB & 16.7 & 7.9 & NB & 33.3 & 33.6 & \cellcolor[HTML]{EFEFEF}NB & \cellcolor[HTML]{EFEFEF}38.9 & 35.3 \\  
 & clnifarsecsq & MP & 16.7 & 13.9 & \cellcolor[HTML]{EFEFEF}RF & \cellcolor[HTML]{EFEFEF}38.9 & 14.9 & RF & 27.8 & 12.4 & RF & 33.3 & 15.6 \\  
 & clnifarsectwo & MP & 11.1 & 7.7 & NB & 61.1 & 50.0 & \cellcolor[HTML]{EFEFEF}NB & \cellcolor[HTML]{EFEFEF}72.2 & 64.9 & NB & 61.1 & 51.9 \\  
\multirow{-8}{*}{Camel} & clnifarsec & LR & 16.7 & 5.0 & NB & 22.2 & 11.6 & LR & 33.3 & 24.9 & \cellcolor[HTML]{EFEFEF}NB & \cellcolor[HTML]{EFEFEF}38.9 & 34.4

\\ 
\multicolumn{14}{r}{~}\\\hline

 & train & NB & 38.1 & 6.8 & NB & 47.6 & 39.3 & NB & 54.7 & 22.2 & \cellcolor[HTML]{EFEFEF}MP & \cellcolor[HTML]{EFEFEF}59.5 & 20.7 \\  
 & farsecsq & KNN & 54.8 & 29.9 & KNN & 59.5 & 40.6 & RF & 54.7 & 51.7 & \cellcolor[HTML]{EFEFEF}RF & \cellcolor[HTML]{EFEFEF}66.7 & 51.5 \\  
 & farsectwo & RF & 47.6 & 12.4 & LR & 59.5 & 24.2 & RF & 47.6 & 27.9 & \cellcolor[HTML]{EFEFEF}MP & \cellcolor[HTML]{EFEFEF}66.7 & 33.6 \\  
 & farsec & NB & 38.1 & 6.3 & NB & 47.6 & 4.1 & NB & 57.1 & 21.0 & \cellcolor[HTML]{EFEFEF}MP & \cellcolor[HTML]{EFEFEF}59.5 & 19.0 \\  
 & clni & RF & 23.8 & 0.4 & NB & 45.2 & 3.5 & NB & 57.7 & 16.8 & \cellcolor[HTML]{EFEFEF}MP & \cellcolor[HTML]{EFEFEF}61.9 & 24.5 \\  
 & clnifarsecsq & KNN & 54.8 & 29.9 & KNN & 59.5 & 42.4 & \cellcolor[HTML]{EFEFEF}RF & \cellcolor[HTML]{EFEFEF}76.2 & 74.7 & RF & 69.0 & 65.1 \\  
 & clnifarsectwo & RF & 35.7 & 9.2 & LR & 59.5 & 24.2 & RF & 54.8 & 36.5 & \cellcolor[HTML]{EFEFEF}LR & \cellcolor[HTML]{EFEFEF}61.9 & 30.3 \\  
\multirow{-8}{*}{Derby} & clnifarsec & NB & 38.1 & 6.8 & NB & 47.6 & 3.9 & \cellcolor[HTML]{EFEFEF}NB & \cellcolor[HTML]{EFEFEF}61.9 & 28.8 & NB & 57.1 & 10.9 \\ 
\end{tabular}

\vspace{8mm} 

\label{tbl:tuningRecall}
\end{table*}

\subsection{Statistics}

This study ranks treatments using the Scott-Knott procedure recommended by Mittas \& Angelis in their 2013 IEEE TSE paper~\cite{Mittas13}. This method
sorts results from different treatments, then splits them in order to maximize the expected value of differences  in the observed performances
before and after divisions. For lists $l,m,n$ of size $\mathit{ls},\mathit{ms},\mathit{ns}$ where $l=m\cup n$, the ``best'' division maximizes $E(\Delta)$; i.e. the difference in the expected mean value before and after the spit: 
 \[E(\Delta)=\frac{ms}{ls}abs(m.\mu - l.\mu)^2 + \frac{ns}{ls}abs(n.\mu - l.\mu)^2\]
Scott-Knott then checks if that ``best'' division is actually useful. To implement that check, Scott-Knott would apply some statistical hypothesis test $H$ to check if $m,n$ are significantly different (and if so, Scott-Knott then recurses on each half of the ``best'' division). For this study, our hypothesis test $H$ was a conjunction of the A12 effect size test of and non-parametric bootstrap sampling; i.e. our Scott-Knott divided the data if {\em both} bootstrapping and an effect size test agreed that
the division was statistically significant (95\% confidence) and not a ``small'' effect ($A12 \ge 0.6$).

For a justification of the use of non-parametric bootstrapping, see Efron \& Tibshirani~\cite[p220-223]{efron93}. For a justification of the use of effect size tests see Kampenes~\cite{kampenes2007} who  warn that even if a hypothesis test declares two populations to be ``significantly'' different, then that result is misleading if the ``effect size'' is very small. Hence, to assess the performance differences  we first must rule out small effects. Vargha and Delaney's non-parametric A12 effect size test 
explores two lists $M$ and $N$ of size $m$ and $n$:

\[
A12 = \left(\sum_{x\in M, y \in N} 
\begin{cases} 
1   & \mathit{if}\; x > y\\
0.5 & \mathit{if}\; x == y
\end{cases}\right) / (mn)
\]

This expression computes the probability that numbers in one sample are bigger than in another. This test was endorsed by Arcuri and Briand~\cite{arcuri2011}.

Table~\ref{tbl:tuningRecall}  presents the reports of our Scott-Knott procedure for each project data set.  These results are discussed, extensively,
over the next two pages. Before that, we address one detail. The ``Learner'' column of that table shows what data miner was found to give the best results for a particular data set/filter combination. Overall, there is no single best learner. That is, when commissioning a vulnerability predictor in a new context, it is important to try a range of different data mining algorithms. 

That said, we note from Table~\ref{tbl:tuningRecall}  that nearest neighbor methods and multi-layer perceptrons (KNN and MLP) appear less often in this table than Bayesian, regression,
or tree-based learners (NB, LR, RF). That is, if engineers wanted to restrict the space of learners they explore, then, 
 they could elect to ignore 
KNN and MLP.

\section{Results}
\label{results}

We now answer our research questions:


\begin{RQ}
{\bf RQ1.} Can learner hyperparameter optimization technique improves security bug report classification performance?
\end{RQ}






\begin{table*}[!b]
\normalsize
\centering
\caption{Average runtime (in minute) of tuning different learners' hyperparameters and SMOTE's hyperparameters.}
\begin{tabular}{c|r|r|r|r|r|r|r|r|r|r|r|r|r|r|r}
\hline
\rowcolor[HTML]{EFEFEF} 
\cellcolor[HTML]{EFEFEF} & \multicolumn{5}{c|}{\cellcolor[HTML]{EFEFEF}\textbf{DE3}} & \multicolumn{5}{c|}{\cellcolor[HTML]{EFEFEF}\textbf{DE10}} & \multicolumn{5}{c}{\cellcolor[HTML]{EFEFEF}\textbf{SMOTUNED}} \\ \cline{2-16} 
\rowcolor[HTML]{EFEFEF} 
\multirow{-2}{*}{\cellcolor[HTML]{EFEFEF}\textbf{Project}} & \multicolumn{1}{c|}{\cellcolor[HTML]{EFEFEF}\textbf{NB}} & \multicolumn{1}{c|}{\cellcolor[HTML]{EFEFEF}\textbf{RF}} & \multicolumn{1}{c|}{\cellcolor[HTML]{EFEFEF}\textbf{MP}} & \multicolumn{1}{c|}{\cellcolor[HTML]{EFEFEF}\textbf{LR}} & \multicolumn{1}{c|}{\cellcolor[HTML]{EFEFEF}\textbf{KNN}} & \multicolumn{1}{c|}{\cellcolor[HTML]{EFEFEF}\textbf{NB}} & \multicolumn{1}{c|}{\cellcolor[HTML]{EFEFEF}\textbf{RF}} & \multicolumn{1}{c|}{\cellcolor[HTML]{EFEFEF}\textbf{MP}} & \multicolumn{1}{c|}{\cellcolor[HTML]{EFEFEF}\textbf{LR}} & \multicolumn{1}{c|}{\cellcolor[HTML]{EFEFEF}\textbf{KNN}} & \multicolumn{1}{c|}{\cellcolor[HTML]{EFEFEF}\textbf{NB}} & \multicolumn{1}{c|}{\cellcolor[HTML]{EFEFEF}\textbf{RF}} & \multicolumn{1}{c|}{\cellcolor[HTML]{EFEFEF}\textbf{MP}} & \multicolumn{1}{c|}{\cellcolor[HTML]{EFEFEF}\textbf{LR}} & \multicolumn{1}{c}{\cellcolor[HTML]{EFEFEF}\textbf{KNN}} \\ \hline
Chromium & \textless{}1 & 14 & 288 & \textless{}1 & 151 & \textless{}1 & 39 & 435 & 4 & 398 & \textless{}1 & \textless{}1 & 6 & \textless{}1 & 11 \\
Wicket & \textless{}1 & \textless{}1 & 4 & \textless{}1 & \textless{}1 & \textless{}1 & 2 & 6 & \textless{}1 & \textless{}1 & \textless{}1 & \textless{}1 & 4 & \textless{}1 & \textless{}1 \\ 
Ambari & \textless{}1 & \textless{}1 & 4 & \textless{}1 & \textless{}1 & \textless{}1 & 2 & 6 & \textless{}1 & \textless{}1 & \textless{}1 & \textless{}1 & 4 & \textless{}1 & \textless{}1 \\ 
Camel & \textless{}1 & \textless{}1 & 4 & \textless{}1 & \textless{}1 & \textless{}1 & 2 & 6 & \textless{}1 & \textless{}1 & \textless{}1 & \textless{}1 & 4 & \textless{}1 & \textless{}1 \\
Derby & \textless{}1 & \textless{}1 & 4 & \textless{}1 & \textless{}1 & \textless{}1 & 2 & 6 & \textless{}1 & \textless{}1 & \textless{}1 & \textless{}1 & 4 & \textless{}1 & \textless{}1 \\ \hline
\end{tabular}
\label{tbl:tuningtime}
\end{table*}

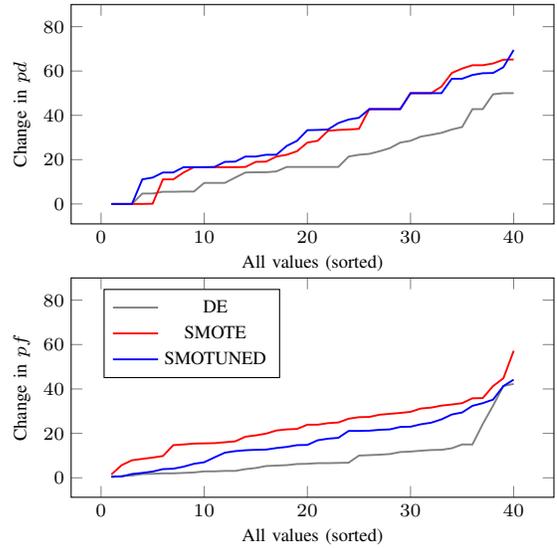
\begin{figure}[!t]
\centering
\scriptsize

  \begin{tikzpicture}
	\begin{axis}[
		height=4.5cm,
		width=8cm,
		ymax=90,
		xlabel={All values (sorted)},
        ylabel={Change in $pd$}
	]
	\addplot[line width=0.25mm, gray] table [x=a, y=DE, col sep=comma, mark=none] {detla_pd.csv};
    \addplot[line width=0.25mm, red] table [x=a, y=SMOTE, col sep=comma,mark=none] {detla_pd.csv};
	\addplot[line width=0.25mm, blue] table [x=a, y=SMOTETUNE, col sep=comma,mark=none] {detla_pd.csv};
	\end{axis}
    \end{tikzpicture} 
  
 \begin{tikzpicture}
	\begin{axis}[
		height=4.5cm,
		width=8cm,
		ymax=90,
		xlabel={All values (sorted)},
        ylabel={Change in $pf$},   
        legend style={at={(0.25,0.95)},anchor=north} 
	]
	\addplot[line width=0.25mm, gray] table [x=a, y=DE, col sep=comma, mark=none] {detla_pf.csv};
	\addlegendentry{DE}
    \addplot[line width=0.25mm, red] table [x=a, y=SMOTE, col sep=comma, mark=none] {detla_pf.csv};
	\addlegendentry{SMOTE}
	\addplot[line width=0.25mm, blue] table [x=a, y=SMOTETUNE, col sep=comma, mark=none] {detla_pf.csv};
	\addlegendentry{SMOTUNED}
	\end{axis}
    \end{tikzpicture}
  
  \caption{Delta between the FARSEC results and the other treatments explored in this paper. The changes in pd and pf
  shown on the y-axis are relative to the the FARSEC results. All these values are positive; i.e. all the new
  treatments of this paper perform much better recall/pd than FARSEC.}\label{ref:ddeltas}
  \end{figure}

Table~\ref{tbl:tuningRecall} reports results
with and without hyperparameter optimization of the pre-processor or learners.
In the table, in each row,
the \colorbox{gray!20}{gray highlights} show all results  ranked ``best'' by a statistical Scott-Knot test. 
Note that:
\bi
\item FARSEC rarely achieves best results.
In fact,
when we compare the hyperparameter optimization technique on learners with FARSEC, we are able to see that the former improves the recall on 33 out of 40 filtering data sets with improvement from 8.6\% to 4.5X.
\item Tuning the learners sometimes produces results that are as good as the best.
\item But most of the best results come from methods
that pre-process the data. Further, tuning the pre-processor (in the SMOTUNED column) most
often leads to the best results.
\ei
Figure~\ref{ref:ddeltas}  visualizes the changes
achieved by the  new methods of this paper.
The changes on y-axis are relative to the   FARSEC results (so a {\em positive} value means
that the treatment is scoring {\em larger} values
than FARSEC). 
 The overall picture offered by Figure~\ref{ref:ddeltas} is that:
 \bi
 \item The ``DE'' treatment (that only tunes the data miners) does not improve  recall (a.k.a. pd) as much as SMOTE/SMOTUNED.
 \item
 The SMOTE/SMOTUNED treatments lead to similar improvements
 in recall;
 \item
 And SMOTUNED do so with a lower false alarm rate.
 \ei 
Several other aspects of Figure~\ref{ref:ddeltas}  
deserve some comments:
\bi
  \item
  The DE results (which come from the {\em Learning Tuning} columns of 
  Table~\ref{tbl:tuningRecall}) have the smallest additional false alarm.
  This  is not particularly
  encouraging, given that the increase in recall/pd seen in the top plot is worse than any other method.
  \item
  The   pf changes resulting from SMOTUNED results are relatively minor compared to the improvements
  seen in recall/pd. Comparing the values in the two plots of Figure~\ref{ref:ddeltas}, we see that
  the median pd improvement of 35\% has a median pf increase of only 18\% (i.e. nearly less than half).
  Also, when going to the maximum values (shown at the right of these plots),
  we see the maximum pd improvement of SMOTUNED is 62\%. This is far greater than the maximum pf increase only 38\%.
  \ei
In summary,  hyperparameter  optimization  technique  improves vulnerability classification performance.
Further, the best results for recall/pd come from tuning the data pre-processor and not the data miner. This comes
at the cost of a relatively minor increase in the false alarm rate.

  \begin{table}[!t]
\caption{A comparison of best result for each project between SMOTE and SMOTUNED.}
\centering
\begin{tabular}{c|c|c|c|c}
\hline
\rowcolor[HTML]{EFEFEF} 
\cellcolor[HTML]{EFEFEF} & \multicolumn{2}{c|}{\cellcolor[HTML]{EFEFEF}\textbf{SMOTE (Best)}} & \multicolumn{2}{c}{\cellcolor[HTML]{EFEFEF}\textbf{SMOTUNED (Best)}} \\ \cline{2-5} 
\rowcolor[HTML]{EFEFEF} 
\multirow{-2}{*}{\cellcolor[HTML]{EFEFEF}\textbf{Project}} & \textbf{pd} & \textbf{pf} & \textbf{pd} & \textbf{pf} \\ \hline
Chromium & 80.8 & 24.1 & 86.1 & 13.6 \\ 
Wicket & 83.3 & 21.1 & 83.3 & 12.1 \\ 
Ambari & 57.1 & 7.9 & 57.1 & 3.9 \\ 
Camel & 72.2 & 12.4 & 61.1 & 15.6 \\ 
Derby & 76.2 & 16.8 & 66.7 & 10.9 \\ \hline
\end{tabular}
\label{tbl:bestCompare}
\end{table}



\begin{RQ}
{\bf RQ2.} Is it best to optimize the learners or the data pre-processors in security bug report classification?
\end{RQ}

The  previous research question showed that off-the-shelf FARSEC could be improved by hyperparameter optimization.
This research question ignores FARSEC and asks if optimizing the learner is better than optimizing the preprocessing.

To answer that question, we look closer that the recall results across the 40 rows on Table~\ref{tbl:tuningRecall}.  Specifically,
we look at the rank that our statistical Scott-Knot test assigns to {\em Learning Tuning}.  If we count how often any
of the SMOTE/SMOTUNED results were ranked worse than {\em Learning Tuning}, we see that:
\bi
\item In  4/40 cases, SMOTE/SMOTUNED is ranked worse than {\em Learning Tuning};
\item In 9/40 cases, SMOTE/SMOTUNED is ranked the same as  {\em Learning Tuning};
\item In the remaining 27/40 cases, SMOTE/SMOTUNED is ranked higher than   {\em Learning Tuning}.
\ei
Hence, we recommend tuning the data pre-processor over tuning the learners.


In our experiment, for each project, we took care to use  the same testing dataset. Therefore, we could pick and compare the best filtering result from SMOTE and SMOTUNED for each data set.
 Table~\ref{tbl:bestCompare} shows the best method (with highest pd)
 seen for SMOTE and SMOTUNED for across the ten rows of each data set. In three of our five data sets, SMOTUNED performs better than SMOTE.
 Specifically,
 for Chromium, SMOTUNED has better pd and pf values.
 Also, for Wicket and Ambrai, SMOTUNED has better pf values;
 Hence, given this evidence and Figure~\ref{ref:ddeltas}, we  recommend SMOTUNED over SMOTE (but in the future, when more data is available,
 we should revisit that conclusion).

\begin{figure}[!t]
\centering
\scriptsize
  \begin{tikzpicture}
	\begin{axis}[
		height=4.5cm,
		width=8cm,
		ymax=90,
		xlabel={All values (sorted)},
        ylabel={$pd$},
	]
	\addplot[line width=0.25mm, orange] table [x=id, y=weka_pd, col sep=comma] {sec2.csv};
	\addlegendentry{WEKA}
    \addplot[line width=0.25mm, blue] table [x=id, y=st_pd, col sep=comma, mark=none] {sec2.csv};
	\addlegendentry{SMOTUNED}
	\legend{};
	\end{axis}
    \end{tikzpicture} 
 
  \begin{tikzpicture}
	\begin{axis}[
		height=4.5cm,
		width=8cm,
		ymax=90,
		xlabel={All values (sorted)},
        ylabel={$pf$},
        legend columns=-1,
        legend style={at={(0.35,0.95)},anchor=north}
	]
	\addplot[line width=0.25mm, orange] table [x=id, y=weka_pf, col sep=comma, mark=none, name path=A] {sec2.csv};
	\addlegendentry{FARSEC}
    \addplot[line width=0.25mm, blue] table [x=id, y=st_pf, col sep=comma, mark=none, name path=B] {sec2.csv};
	\addlegendentry{SMOTUNED}
	\end{axis}
    \end{tikzpicture} 
 
\caption{Top row: changes in pd and pf values among DE, SMOTE and SMOTUNED (compared to FARSEC). Bottom row: 
A comparison of pd and pf values between FARSEC and SMOTUNED.}
\label{fig:comparison}
\end{figure}
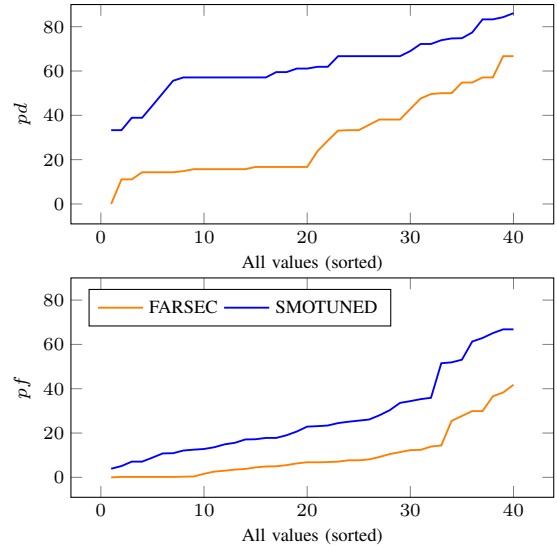

 Figure~\ref{fig:comparison}  visualizes the overall improvements  in pf and  recall  (a.k.a.  pd)  achieved
by SMOTUNED.
This plot shows the {\em absolute} performance  the  best  methods  of  this  paper (SMOTUNED) compared to the prior state of the art (FARSEC).
Note that the median performance scores for FARSEC and not so impressive (pd=20\%)
while the SMOTUNED recalls are much better (pd=63\%). These improvements come at relatively minor cost
(median false alarm rate of 7\% for FARSEC and 18\% for SMOTUNED).  

Two objections to this  endorsement of SMOTUNED for vulnerability classification might be:
\bi
\item 
{\em The runtime cost of tuning.} Hyperparameter optimization can be  a slow process. However, as shown in 
Table~\ref{tbl:tuningtime}, the CPU costs for the methods of this paper are not excessive (exception:
as seen in Table~\ref{tbl:tuningtime},
KNN and MLP can be  very slow to tune. However, since these were rarely selected as a ``best'' learner in Table~\ref{tbl:tuningRecall}, we would recommend not using those learners for vulnerability classification.
\item
{\em The complexity of building and applying the hyperparameter optimizer.} In our experience, it was not hard  to implement a DE algorithm and apply it to vulnerability classification. Indeed, one of the advantages of DEs is that they are {\em very} simple to build. Nevertheless, for those that want to avoid that (very small) additional
complexity, we offer them the advice from our answer to {\bf RQ3}.
\ei

\begin{RQ}
{\bf RQ3.} What are the relative merits of irrelevancy pruning (e.g., filtering) vs hyperparamater optimization for security bug report classification for security bug report classification?
\end{RQ}

Our results show that with data pre-processing optimization, there is no added  benefit to FARCEC’s irrelevancy pruning.
That said, the {\em train} pd results in the {\em FARSEC} column of Table~\ref{tbl:tuningRecall} are usually smaller than the other values seen after applying
some of the filters proposed by Peters et al. 
This result confirms the Peters et al. results;
i,e, without hyperparameter optimization, irrelevancy pruning does indeed improve the performance of vulnerability classification. 

Hence,  while we recommend data pre-processing optimization, if analysts wish to avoid that added complexity, they  should use FARSEC’s irrelevancy pruning.

\section{Threats to Validity}
\label{threats}

As to any empirical study, biases can affect the final results. Therefore, conclusions drawn from this work must be considered with threats to validity in mind. In this section, we discuss the validity of our work.

\textbf{Sampling Bias.} Sampling bias threatens any classification experiments, and what matters here may not be true there. For example, the data sets used here come from FARSEC, i.e., one chromium project and four Apache projects in different application domains. Bug report from the chromium project is only a subset of the project, and the bug reports from four Apache projects are generated by JIRA which is used as the bug tracking system. In addition, the bug reports from Apache projects are randomly selected with a BUG or IMPROVEMENT label for each project.

\textbf{Learner Bias.} Classification is a large and active field. Lots of different machine learning algorithms have been developed to solve different classification problem tasks. An important note from David Wolpert is the ``No Free Lunch Theorems'', i.e., there is no one learner working for all problems. One reason for this is that each classification has its inherent biases. Any study can only use a small subset of the known classification algorithms. In this work, we do not explore the performance of different classification learners. To compare the work with FARSEC, we use the same learners as Peters did in their work.

\textbf{Evaluation Bias.} On the one hand, during our evaluation, we repeat our work 10 times and report the median performance results for different learners. On the other hand, we only set two versions of generations (i.e., 3 and 10) for learners and one version of generation (i.e., 10) for SMOTE in differential evolution algorithm. Larger number of generations could be further explored, but with more CPU efforts and time.

\textbf{Input Bias.} In hyperparameter optimization and preprocessing tuning with differential evolution algorithm, we choose a subset of hyperparameters for each learner, and set range boundaries for both learners and SMOTE. Different input values can result in different outputs, we choose these ranges as we explain in Section~\ref{tion:hpo}, and we do not explore large range in this study which is not the focus of the work.

\section{Related Work}
\label{related}

\subsection{Mining Bug Reports}\label{tion:vuln}

Text mining (also known as Text Analytics) is the process of exploring and analyzing massive sets of unstructured (e.g., word documents) or semi-structured (e.g., XML and JSON) text data, in order to identify concepts, patterns, topics and other attributes in the data~\cite{allahyari2017brief}. To pre-process the textual content, several steps such as tokenization, filtering, lemmatization, stop-words removal and stemming are usually leveraged~\cite{hotho2005brief}.

Text mining has recently been widely applied in bug report analysis, such as identification of duplicated bug reports~\cite{sun2011towards,lazar2014improving,hindle2016contextual,deshmukh2017towards}, prediction of the severity or impact of a reported bug~\cite{lamkanfi2010predicting,zhang2015predicting,tian2012information,yang2016automated,yang2017high}, extraction of execution commands and input parameters from performance bug reports~\cite{han2018perflearner}, assignment of the priority labels to bug reports~\cite{tian2015automated}, bug report field reassignment and refinement prediction~\cite{xia2016automated}.

In particular, a few studies of bug report classification are more relevant to our work. Some of those approaches focus on building bug classification models based on analyzing bug reports with text mining. For example, Zhou et al.~\cite{zhou2016combining} leveraged text mining techniques, analyzed the summary parts of bug reports and fed into machine learners. Xia et al~\cite{xia2014automated} developed a framework that applied text mining technology on bug reports, and trained a model on bug reports with known labels (i.e., configuration or non-configuration). The trained model was used to predict the new bug reports. Popstojanova et al.~\cite{goseva2018identification} used different types of textual feature vectors and focused on applying both supervised and unsupervised algorithms in classifying security and non-security related bug reports. Wijayasekara et al.~\cite{wijayasekara2014vulnerability} extracted textual information by utilizing the textual description of the bug reports. A feature vector was generated through the textual information, and then presented to a machine learning classifier.

Some other approaches use a more heuristic way to identify bug reports. For example, Zaman et al.~\cite{zaman2011security} combined keyword searching and statistical sampling to distinguish between performance bugs and security bugs in Firefox bug reports. Gegick et al.~\cite{gegick2010identifying} proposed a technique to identify security bug reports based on keyword mining, and performed an empirical study based on an industry bug repository.


\subsection{Pre-processor Optimization}

Agrawal et al.~\cite{agrawal2018better} report similar results to those seen in {\bf RQ2}. They also report that data pre-processor optimization does better than leaner optimization -- which they summarize as ``better data'' does better than ``better learners''.

While Argawal et al. might have predicted our {\bf RQ2} result, there are many reasons to believe that such a prediction
could be incorrect. Argawal et al. only studied defect classification and methods that work for defect classification usually do not work for vulnerability classification. The historical record is very clear on this point. As shown in \tion{vuln}, when defect classification methods are applied to vulnerability classification, they do not perform very well. This is because the kinds of data explored by defect classification and vulnerability classification are very different. The vulnerability frequencies reported in the Introduction (1-4\% of files) are very common. Yet defect classification usually is performed on data with much larger target classes.
For example, the defect data in a recent defect classification modeling (DPM) paper used 11 data sets with median number of files with defects of 34\%~\cite{krishna2017less}.  

Hence we say that prior to this paper, it was an open question if methods (like those of Agrawal et al.) which are certified for one kind of software engineering problem (defect classification) work for a very different kind of problem (vulnerability classification).

\section{Conclusion}
\label{conclusion}

Security vulnerabilities are a pressing problem
that threatens not only the viability of software services, but also consumer confidence in those services. Prior works in security bug report classification have had issues with the scarcity of vulnerability data (specifically, such incidents occur very rarely). In a recent TSE'18 paper, Peters et al. proposed some novel filtering algorithms to help improve security bug report classification. This paper reproduced those FARSEC results to show that, yes indeed, such filtering can improve bug report classification.

Our experiments also show that we can do better than FARSEC using hyperparameter optimization of learners and data pre-processors. Of those two approaches, our results show that it is more advantageous to improve the data-processor than the learner. Hence, in the future, we strongly recommend the use of hyperparameter optimization for improving data pre-processing for software bug report classification. 

In closing, we feel it important to repeat and emphasize some advice offered earlier in this paper. Data mining algorithms should be assessed on the kinds of data they might see in the future. That is, data miners should be tested on data that has {\em naturally occurring class distributions}. Hence, while applying SMOTUNED (or SMOTE) to the {\em training data} is good practice, it is a bad practice to apply such pre-processing to the {\em testing data}.

\balance
\bibliographystyle{plain}

\end{document}